\documentclass[aps,prl,twocolumn,superscriptaddress]{revtex4}

\usepackage{graphicx}
\usepackage{amssymb}
\usepackage{xcolor}

\begin{document}

\title{Dispersing Nanoparticles in a Polymer Film via Solvent Evaporation}

\author{Shengfeng Cheng}
\email{chengsf@vt.edu}
\affiliation{Department of Physics, Center for Soft Matter and Biological Physics, and Macromolecules Innovation Institute,
Virginia Polytechnic Institute and State University,\\
Blacksburg, Virginia 24061, USA}
\author{Gary S. Grest}
\affiliation{Sandia National Laboratories, 
Albuquerque, NM 87185, USA}

\date{\today}

\begin{abstract}
Large scale molecular dynamics simulations are used to study the dispersion of nanoparticles (NPs) in a polymer film during solvent evaporation. As the solvent evaporates, a dense polymer-rich skin layer forms at the liquid/vapor interface, which is either NP rich or poor depending on the strength of the NP/polymer interaction. When the NPs are strongly wet by the polymer, the NPs accumulate at the interface and form layers. However when the NPs are only partially wet by the polymer, most NPs are uniformly distributed in the bulk of the polymer film with the dense skin layer serving as a barrier to prevent the NPs from moving to the interface. Our results point to a possible route to employ less favorable NP/polymer interactions and fast solvent evaporation to uniformly disperse NPs in a polymer film, contrary to the common belief that strong NP/polymer attractions are needed to make NPs well dispersed in polymer nanocomposites.
\end{abstract}

\maketitle

A polymer nanocomposite (PNC) consists of a polymer matrix in which nanofillers (e.g., nanoparticles, nanorods, nanofibers, nanotubes, etc.) are embedded. PNCs have recently attracted significant attention because of their increasingly wide range of potential applications resulting from the fact that the addition of nanofillers leads to improved properties.\cite{hussain06,kumar10} 
Previous studies have established that the quality of nanofiller dispersion in the polymer matrix and the nanofiller/polymer interface play dominant roles in controlling the properties of PNCs.\cite{mackay06,jancar10,jouault12} Various strategies have been developed to control the dispersion of nanofillers by delicately balancing equilibrium factors including energetic interactions and entropic effects.\cite{mackay06,krishnan07,jouault14b} However, the manufacturing process typically involves procedures that are intrinsically out-of-equilibrium and it is not clear how processing affects the distribution of nanofillers in PNCs. Jouault et al. pointed out that processing is critical in determining the initial nanofiller dispersion state and in many cases subsequently annealing does not alter this state significantly.\cite{jouault14b} This observation indicates that it may be extremely difficult for the distribution of nanofillers in a polymer host to reach thermodynamic equilibrium.

One frequently employed method to fabricate PNCs is solvent casting: polymers and nanofillers are first dispersed in a solvent (or a mixture of solvents) and the solvent is evaporated.\cite{mbhele03,maximous09,vyostskii11,jouault14b,imel15,beg15} Previous work by Jouault et al. showed that using different casting solvents can lead to either dispersion or aggregation of the same nanoparticles (NPs) in the same polymer matrix.\cite{jouault14b} The nonequilibrium nature of the evaporation process is further expected to influence the distribution of nanofillers. However, such a seemingly important issue remains largely unexplored for PNCs, though in the context of evaporation-induced self-assembly of NPs the evaporation rate has been shown to have a strong effect on the assembly structures.\cite{bigioni06,ojha10,cheng13JCP} Previous work showed that the evaporation rate is a critical factor in the drying of polymer films and paint.\cite{composto90,strawnhecker01,koombhongse01,luo05,erkselius08,zhang12,kooij15} There has been some evidence that the NP dispersion can be improved by evaporating the solvent quickly.\cite{sen07,hu12} However, it is still unclear how the complex interplay of energetics, entropy, kinetics, and evaporation conditions controls the nanofiller dispersion and the properties of PNCs.\cite{jouault14b} Understanding the role of these factors may yield fresh insights on new strategies to control the dispersion state of nanofillers. Here we report the results from large-scale molecular dynamics (MD) simulations of the effect of solvent evaporation on the dispersion of NPs in a polymer film. We show that the nonequilibrium nature of the evaporation process, coupled with tuned NP/polymer interactions, can dramatically affect NP dispersion in unexpected ways. We identify a possible mechanism to uniformly disperse NPs in thin polymer films when the NP/polymer interaction favors phase separation and propose ways to experimentally test the mechanism.

\begin{figure}[ht]
\centering
\includegraphics[width=3.25in]{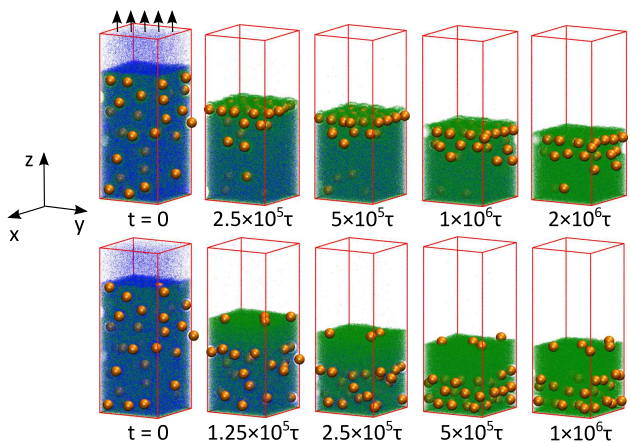}
\caption{Temporal evolution of the NP distribution as the solvent evaporates. Top row: $A_{\rm np}=200\epsilon$; bottom row: $A_{\rm np}=80\epsilon$. Only about a quarter of the system in the $x$-$y$ plane is shown; part of the vapor region is also cut for a better visualization. Color scheme: NP (orange), polymer (green), and solvent (blue).}
\label{assembly}
\end{figure}

\begin{figure}[ht]
\centering
\includegraphics[width=3in]{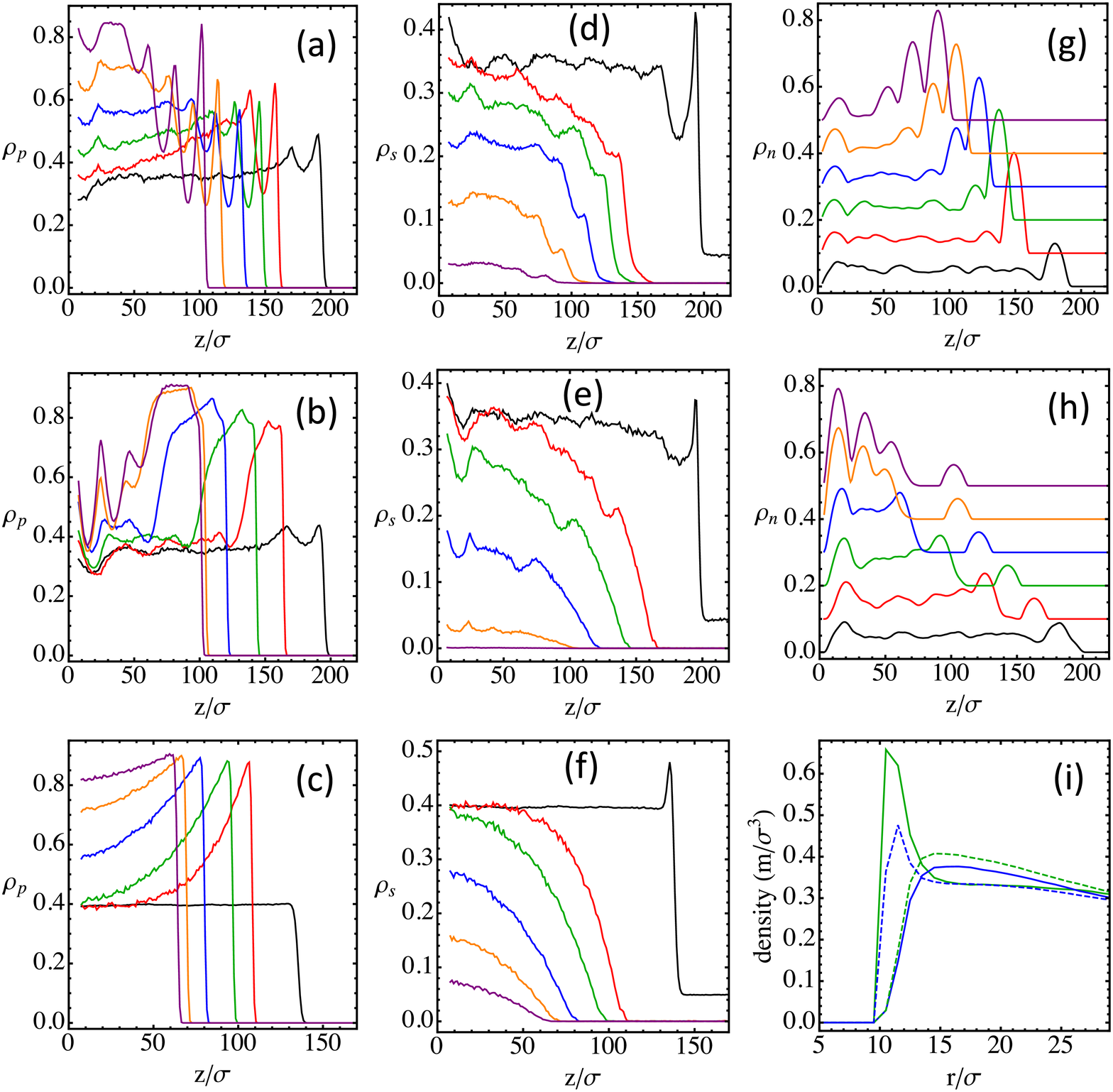}
\caption{Density profiles at a series of times during evaporation for the polymer [(a)-(c)], solvent [(d)-(f)], and NPs [(g) and (h)]. Plots (a), (d), and (g) are for $A_{\rm np}=200\epsilon$ at $10^{-6}t/\tau=0$ (black), $0.125$ (red), $0.25$ (green), $0.5$ (blue), $1.0$ (orange), and $2.0$ (purple). Plots (b), (e), and (h) are for $A_{\rm np}=80\epsilon$ at $10^{-6}t/\tau=0$ (black), $0.05$ (red), $0.125$ (green), $0.25$ (blue), $0.5$ (orange), and $1.0$ (purple). For clarity each profile in (g) and (h) is shifted upward by $0.1\sigma^{-3}$ from the earlier one. Plots (c) and (f) are for the neat polymer solution at $10^{-5}t/\tau=0$ (black), $0.25$ (red), $0.5$ (green), $1.0$ (blue), $1.5$ (orange), and $2.0$ (purple). Plot (i) is the distribution of solvent (blue) and polymer beads (green) around an NP prior to evaporation of the solvent for $A_{\rm np}=200\epsilon$ (solid lines) and $A_{\rm np}=80\epsilon$ (dashed lines).}
\label{density}
\end{figure}

We modeled a system of 29,217 linear polymer chains of length 100.\cite{kremer90} The solvent consists of 2.92 million single beads identical to the polymer monomers. All beads have mass $m$ and interact through a standard Lennard-Jones (LJ) 12-6 pairwise potential with strength $\epsilon$ and characteristic length $\sigma$. The time unit is $\tau \equiv\sqrt{m\sigma^2/\epsilon}$. The system contains 200 NPs of diameter $d=20\sigma$. The interaction between NPs is given by an integrated LJ potential characterized by the Hamaker constant $A_{\rm nn} = 39.48\epsilon$.\cite{Everaers03} The interaction between NPs is chosen to be purely repulsive, corresponding physically to adding a short surfactant coating on NPs to avoid flocculation.\cite{veld09,grest11} The interaction between an NP and a solvent or polymer bead is determined by a similar integrated potential with Hamaker constants $A_{\rm ns}$ and $A_{\rm np}$, respectively. We set $A_{\rm ns}=100\epsilon$, in which case the NPs would be fully solvated in the solvent if no polymer was present.\cite{cheng12JCP} We have compared two NP systems: one has a stronger NP/polymer interaction with $A_{\rm np}=200\epsilon$; while the other has a weaker NP/polymer interaction with $A_{\rm np}=80\epsilon$. In the former case the NPs are strongly wet by the polymer while in the latter case, the NPs are only partially wet by the polymer. For $A_{\rm np}=80\epsilon$ a NP at the liquid/vapor interface of a melt of chain length 100 has a contact angle $\sim 96^\circ$, while for $A_{\rm np}=200\epsilon$ the NPs placed at the interface diffuse into the polymer matrix\cite{cheng12JCP,meng13SM} and each NP is coated with a bound polymer layer [see Fig.~\ref{density}(i)] which helps enhance the NP dispersion.\cite{jouault14b} For comparison, we have also simulated the evaporation process for a neat polymer solution without NPs but with all the other interactions kept the same. The NP systems are shown in Fig.~\ref{assembly}. After the systems are equilibrated, we add a deletion zone of thickness $20\sigma$ above the vapor. Every $0.5\tau$ all the solvent beads in the deletion zone are removed, effectively mimicking fast solvent evaporation into a vacuum.\cite{cheng11} All the simulation details are given in the Supporting Information.

For all three systems, a polymer-rich skin layer quickly forms at the liquid/vapor interface as the solvent evaporates (Fig.~\ref{assembly}). A concentrated polymer layer at the interface was previously observed experimentally during the evaporation of a polymer solution \cite{composto90,haas00,strawnhecker01,koombhongse01,wang13} and was studied with phenomenological models \cite{okuzono06,muench11} and simulations \cite{tsige04a,tsige04b,tsige05}. The formation of the skin layer induces density gradients (with opposite signs) for the solvent and polymer near the interface and has a strong effect on the distribution of NPs in the resulting polymer film. For $A_{\rm np}=200\epsilon$, the NP/polymer interaction is stronger than the NP/solvent interaction, which is reflected in the distribution of polymer and solvent beads around an NP, as shown in Fig.~\ref{density}(i). The surface polymer layer thus provides a more favored solvation environment for the NPs, which start to accumulate in this skin layer. As the evaporation proceeds, the surface polymer layer thickens and entraps more and more NPs, which start to form a well-organized layer. After the first layer of NPs is nearly complete, a second layer of NPs starts to form just below the first layer. The process continues with the formation of multilayers of NPs with the number of layers dependent on the NP concentration. In Fig.~\ref{assembly} three layers are observed when almost all ($\sim 97\%$) the solvent has evaporated. Note that $T=1.0\epsilon/k_{\rm B} > T_g$ (the glass transition temperature of the polymer film). For $A_{\rm np}=200\epsilon$ as the NPs are completely wet by the polymer, given enough time the NPs are expected to diffuse back to the polymer film, leading to a more uniform NP distribution. To prevent this from occurring the temperature would need to be quenched to below $T_g$ before the NPs have time to diffuse into the film. Alternatively one can use a polymer that is plasticized by the solvent so that once the solvent evaporates, the polymer film is below $T_g$, thereby inhibiting the diffusion of the NPs back into the film.

When the polymer only partially wets the NP ($A_{\rm np}=80\epsilon$) we see a dramatically different behavior as the solvent evaporates. In this case, the NPs are almost excluded from the polymer-enhanced surface layer since the NP/polymer interaction is less favorable compared to the NP/solvent interaction [see Fig.~\ref{density}(i)]. While a few NPs that are initially close to the liquid/vapor interface move to the surface of the film, most of the NPs are dispersed uniformly in the film. Only in the final stage of evaporation do the NPs start to form layers as the total thickness of the film is reduced, as shown in Fig.~\ref{assembly}. As in the case with $A_{\rm ns}=200\epsilon$, to retain this state of NP distribution one would need to quench the system below $T_g$ before the NPs have time to diffuse or use a polymer that is plasticized by the solvent.

To quantify the distribution of the polymer, solvent, and NPs we plot in Fig.~\ref{density} the density profiles of all three components at various times. The density is defined as $\rho_i (z)=n_i(z)m_i/(L_x L_y \Delta z)$ where $n_i(z)$ designates the number of $i$-type particles in the spatial bin $[z-\Delta z/2, z+\Delta z/2]$. Since we set $\Delta z = 1.0\sigma$, a NP with $20\sigma$ diameter straddles several bins and we partition the NP mass to bins based on the partial volume of the NP enclosed by each bin, i.e., the contribution to the corresponding $n_i(z)$ from each NP is a fraction. The solvent and polymer beads are treated as a point mass in the calculation of $\rho_i(z)$. Subtracting the volume occupied by the NPs in each spatial bin when the solvent or polymer density is calculated only leads to minor changes in the results shown in Fig.~\ref{density}.

As the evaporation proceeds, the thickness of the film is reduced and the liquid/vapor interface moves towards the lower wall at $z=0$, as shown in Fig.~\ref{density}. The formation of the surface polymer layer is observed in all cases.\cite{tsige04a,tsige04b,tsige05} The solvent density decreases with $z$ and shows a parabola-like profile, which is most obvious for the case of a neat polymer solution (Fig.~\ref{density}(f)). For $A_{\rm np}=200\epsilon$, as the NPs are accumulated in the polymer-rich skin layer, alternating density peaks of NPs and polymer chains are observed (Fig.~\ref{density}(a) and (g)) and the NPs are depleted in the region below the surface polymer layer. For $A_{\rm np}=80\epsilon$, the NPs are excluded from the polymer skin layer and dispersed in the region below the surface layer (Fig.~\ref{density}(h)). After most of the solvent is evaporated, confinement causes the NPs to form layers near the lower wall. Similar results were observed for systems with initially lower NP and polymer volume fractions (see Fig. S1 in the Supporting Information).

\begin{figure}[ht]
\centering
\includegraphics[width=3.25in]{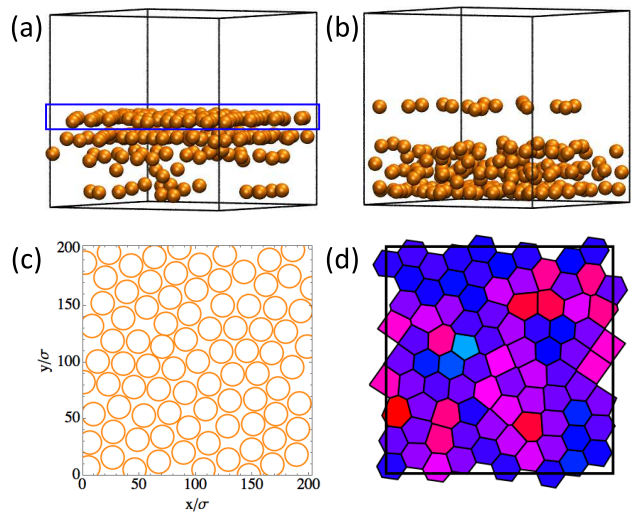}
\caption{(a) and (b): Snapshots of NP dispersion in the polymer film after almost all ($97\%$) the solvent is evaporated: (a) $A_{\rm np}=200\epsilon$ and (b) $A_{\rm np}=80\epsilon$. (c) and (d): the structure in the first layer of NPs that are enclosed by the blue box in (a); (c) shows the NPs as circles and (d) is the corresponding Voronoi reconstruction.}
\label{top_layer}
\end{figure}

The accumulation and layering of NPs in the surface polymer layer for $A_{\rm np}=200\epsilon$ are clearly seen in Fig.~\ref{top_layer}(a), where the final distribution of NPs is shown after almost all ($\sim 97\%$) the solvent is evaporated. Three layers of NPs are observed, corresponding to the three density peaks in Fig.~\ref{density}(g). Note that some NPs stay close to the lower wall because of the attractive interaction with the wall. In Figs.~\ref{top_layer}(c) and (d), the organization in the first layer of NPs and the corresponding Voronoi construction are shown. Overall, the NPs tend form a close-packed hexagonal structure. However, square as well as face-centered cubic packing and many defects are also observed. The diffusion of NPs is too slow to remove the defects and yield a closed-packed two-dimensional lattice on the time scale of our simulations. Fig.~\ref{top_layer}(b) shows the distribution of all NPs in the system with $A_{\rm np}=80\epsilon$ after the solvent evaporation, where the expelling of NPs from the polymer-rich skin layer at the liquid/vapor interface is obvious. The dispersion of NPs in the region below this layer is almost uniform.

\begin{figure}[ht]
\centering
\includegraphics[width=3in]{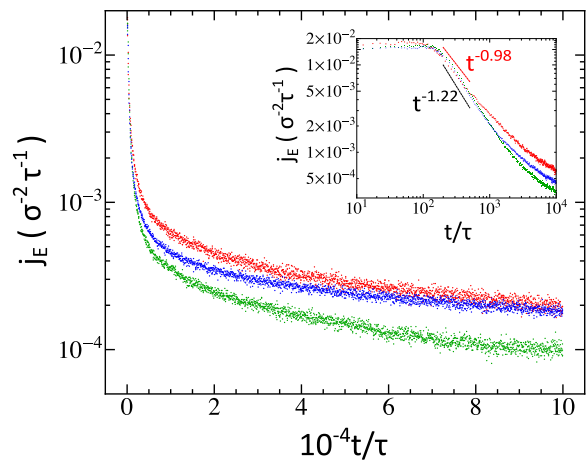}
\caption{Log-linear plot of evaporation rate $j_E$ as a function of time ($t$) for the neat polymer solution (red), the NP system with $A_{\rm np}=200\epsilon$ (green), and the NP system with $A_{\rm np}=80\epsilon$ (blue). Inset: log-log plot of $j_E$ vs. $t$ for short time.}
\label{evap_rate}
\end{figure}

Figure~\ref{evap_rate} shows the evaporation rate $j_E$ as a function time ($t$) for all three systems we have simulated. The rate is defined as $j_E = (1/L_xL_y)\times{\rm d}N/{\rm d}t$ with $N$ as the number of solvent beads remaining in the simulation box; $j_E$ represents the number of solvent beads entering the deletion zone per unit area per unit time. For all three systems, $j_E$ is high and remains roughly constant for $t \lesssim 100\tau$ as the solvent in the vapor phase is evaporated. The time scale of this regime thus depends on the thickness of the vapor phase (i.e., the distance between the liquid/vapor interface and the deletion zone). For $t \gtrsim 100\tau$ the evaporation rate decreases with time as the polymer rich layer near the interface forms and inhibits the solvent diffusion to the surface. In the early stage ($200\tau\lesssim t\lesssim 500\tau$) $j_E \sim t^{-\alpha}$ with $\alpha\approx 0.98$ for the neat solution and $1.22$ for the NP systems. For $t\gtrsim 500\tau$ the neat solution has an evaporation rate higher than the NP systems because in the former no NPs are present at the liquid/vapor interface to block the evaporation. For the NP systems, the NP/polymer interaction also has an effect on $j_E$. At early time, the two NP systems exhibit similar evaporation rates. However, for $t\gtrsim 1000\tau$ the rate $j_E$ decreases faster for $A_{\rm np}=200\epsilon$ since the NPs accumulate in the surface polymer layer, reducing the effective area for evaporation. For $A_{\rm np}=80\epsilon$, the NPs are excluded from the surface layer so the evaporation rate is similar to that of the neat solution at late time, as shown in Fig.~\ref{evap_rate}.

\begin{figure}[ht]
\centering
\includegraphics[width=3.25in]{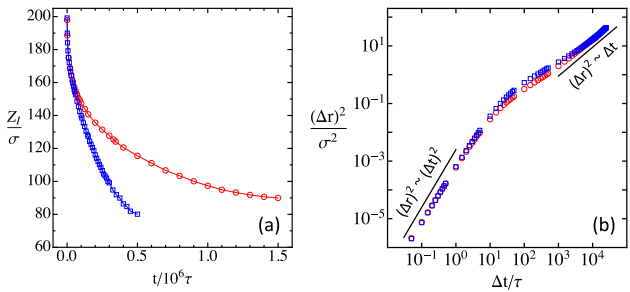}
\caption{(a) Location of the liquid-vapor interface ($Z_I$) vs. time; (b) mean square displacement of NPs vs. time in the polymer/solvent mixture before evaporation of the solvent for $A_{\rm np}=200\epsilon$ (circles) and $A_{\rm np}=80\epsilon$ (squares).}
\label{peclet}
\end{figure}

In the context of drying, a dimensionless P\'{e}clet number can be defined to describe the competition between the time scale of diffusion and that of evaporation.\cite{trueman12,fortini16} Here the diffusion time scale is $\tau_{\rm D} = H^2/D$ where $H$ is the film thickness and $D$ the diffusion coefficient of NPs. The evaporation time scale can be roughly defined as $\tau_{\rm E} = H/v_s$ where $v_s$ is the descending speed of the liquid/vapor interface or the rate at which the film thickness decreases. The P\'{e}clet number is $Pe = \tau_{\rm D}/\tau_{\rm E} = Hv_s/D$. To estimate $v_s$, we calculated the location $Z_I$ of the liquid/vapor interface as a function of time, which as shown in Fig.~\ref{peclet}(a) is a non-linear function of time. Although $v_s$ is usually assumed to be constant in mesoscopic models,\cite{fortini16} $v_s \equiv dZ_I/dt$ clearly decreases with increasing time. For simplicity, we take the average descending speed of the interface as $\overline{v}_s$, which is $\sim 7\times 10^{-5}\sigma/\tau$ for $A_{\rm np}=200\epsilon$ and $\sim 2\times 10^{-4}\sigma/\tau$ for $A_{\rm np}=80\epsilon$.

The diffusive motion of NPs during solvent evaporation is difficult to quantify since the film thickness is reduced with time, which imposes a convective component on the NP motion. To estimate $Pe$ we use the diffusion coefficient of NPs in the equilibrium polymer solution before evaporation. The equilibrium mean square displacement of NPs is shown in Fig.~\ref{peclet}(b), which shows the typical ballistic motion at short time and diffusive behavior at long time. For both systems $D \approx 3\times 10^{-4}\sigma^2/\tau$. Combining the estimated values of $D$, $\overline{v}_s$, and $H \approx 100\sigma$, we find that $Pe=H\overline{v}_s/D$ is $\sim 20$ for $A_{\rm np}=200\epsilon$ and $\sim 70$ for $A_{\rm np}=80\epsilon$. As shown in Figs.~\ref{assembly} and \ref{density} the distribution of NPs in the polymer film is largely set in the early stage of evaporation when the instantaneous value of $v_s$ is larger than $\overline{v}_s$. During solvent evaporation, the diffusion of NPs slows down and the diffusion coefficient decreases from the equilibrium value as the evaporation proceeds. As a result, $Pe$ is even larger in the important, early stage of evaporation.

In the limit $Pe \gg 1$, the interface moves much faster than the NPs diffuse. The moving interface impinges on the NPs, which cannot diffuse quickly enough back to the film. As a result, the NPs accumulate near the descending interface at the top of the film.\cite{fortini16} Our results for $A_{\rm np}=200\epsilon$ are consistent with this prediction; the accumulation of NPs is further enhanced by the favored NP/polymer interaction. While $Pe$ is even larger for the system with $A_{\rm np}=80\epsilon$, the formation of the surface polymer layer at the liquid/vapor interface impedes the transport of NPs to the interface. In this case the NPs accumulate slightly below the surface polymer layer, reminiscent of the expected behavior at large $Pe$.

To achieve large $Pe$ experimentally, the solvent must evaporate at a fast rate such that $v_s \gg D/H$. The effect of evaporation rate has long been noticed on the morphology of polymer films.\cite{composto90,strawnhecker01,koombhongse01,luo05,erkselius08,zhang12,kooij15} Recently, driving particle assembly with fast evaporation has been demonstrated for gold NPs using a one-step, near-infrared radiation-assisted evaporation process.\cite{utgenannt16}. It is interesting to test experimentally if fast evaporation can be combined with tuned NP/polymer interactions to control the NP dispersion in polymer films. That NPs move to the surface of a polymer film during solvent evaporation has been frequently observed in experiments where the NPs and polymers are immiscible.\cite{meli08,meli09,green11} Our simulations show that this can also occur with miscible NPs when the solvent evaporates rapidly. We should note that the conditions here are different from those in the experiment by Krishnan et al.\cite{krishnan07} who showed that the NPs diffuse to the surface of either the substrate or the polymer film after thermal annealing. In our case the accumulation (depletion) of NPs near the film (substrate) surface is induced by the fast evaporation that magnifies the ``nonequilibrium'' factor. Our results further show that for the immiscible NP/polymer systems, the NPs can still remain dispersed in the polymer film after solvent evaporation. The key strategy is to use fast evaporation to drive the quick formation of the surface polymer layer, which serves as a barrier to prevent the NPs from diffusing to the surface of the film.

\section*{Acknowledgments}
This research used resources of the National Energy Research Scientific Computing Center (NERSC), which is supported by the Office of Science of the United States Department of Energy under Contract No. DE-AC02-05CH11231. These resources were obtained through the Advanced Scientific Computing Research (ASCR) Leadership Computing Challenge (ALCC). This work was performed, in part, at the Center for Integrated Nanotechnologies, an Office of Science User Facility operated for the U.S. Department of Energy (DOE) Office of Science. Sandia National Laboratories is a multi-program laboratory managed and operated by Sandia Corporation, a wholly owned subsidiary of Lockheed Martin Corporation, for the U.S. Department of Energy's National Nuclear Security Administration under contract DE-AC04-94AL85000.


\section{Supporting Information for ``Dispersing Nanoparticles in a Polymer Film via Solvent Evaporation"}

\section{Simulation Methods}
We modeled the polymer as bead-spring linear chains. Each chain consists of 100 coarse-grained beads. The solvent consists of single beads that are the same as the polymer monomers. All polymer and solvent beads have mass $m$ and interact through a standard Lennard-Jones (LJ) 12-6 pairwise potential, $U_{\rm LJ}(r)=4\epsilon [(\sigma/r)^{12}-(\sigma/r)^6 -(\sigma/r_c)^{12}+(\sigma/r_c)^6 ]$, where $r$ is the distance between the centers of two beads, $\epsilon$ the energy scale and the strength of interaction, and $\sigma$ the size of beads. The cut off distance is $r_c=3.0\sigma$ for all non-bonded pairs and $r_c=2^{1/6}\sigma$ for bonded pairs of neighboring polymer beads on a chain. The latter are connected by an additional bond given by the finitely extensible nonlinear elastic (FENE) potential $U_{B}(r) = -\frac{1}{2} K R_0^2 {\rm ln}[ 1-( r/R_0 )^2 ]$, where $r$ is the bond length and $R_0=1.5\sigma$ and $K=30\epsilon/\sigma^2$.\cite{kremer90}

The NPs are modeled as spheres of diameter $d=20\sigma$. Each NP is treated as a uniform distribution of LJ particles of size $\sigma$ and at density $1.0m/\sigma^3$, resulting in a NP mass $M=4188.8m$. The interaction between NPs can be determined analytically by integrating over all the interacting pairs between the two NPs.\cite{Everaers03} The strength of the resulting potential is characterized by the Hamaker constant $A_{\rm nn} = 39.48\epsilon$. For simplicity, the interaction between NPs is chosen to be purely repulsive with the cut off set at $20.427\sigma$. This corresponds physically to adding a short surfactant coating on NPs to avoid flocculation.\cite{veld09,grest11} NP/NP attractions can also be added but are not expected to affect the results reported here as long as NPs do not aggregate in the polymer solution before the solvent evaporates. The interaction between an NP and a solvent or polymer bead is determined by a similar integrated potential with Hamaker constants $A_{\rm ns}$ and $A_{\rm np}$, respectively; both the NP/polymer and NP/solvent interactions are truncated at $14\sigma$. We set $A_{\rm ns}=100\epsilon$, in which case the NPs would be fully solvated in the solvent if no polymer was present.\cite{cheng12JCP} We have compared two NP systems: one has a stronger NP/polymer interaction with $A_{\rm np}=200\epsilon$; while the other has a weaker NP/polymer interaction with $A_{\rm np}=80\epsilon$.

The simulation cell is a rectangular box of dimensions $L_x \times L_y \times L_z$. The liquid/vapor interface is parallel to the $x$-$y$ plane, in which periodic boundary conditions are imposed. In the $z$ direction, all the particles are confined between two flat walls at $z=0$ and $z=L_z$, respectively. The wall/particle interactions are represented with a LJ 9-3 potential $U_{\rm W}(h)=4\epsilon_W [(2/15)(D/h)^{9}-(D/h)^3 -(2/15)(D/h_c)^{9}+(D/h_c)^3 ]$, where $\epsilon_W$ is the interaction strength, $D$ the characteristic length, $h$ the separation between the center of a particle and the wall, and $h_c$ the cut off separation. At the lower wall, we set $\epsilon_W=1.0\epsilon$ ($4.0\epsilon$), $D=1.0\sigma$, $h_c=3.0\sigma$ for the solvent/wall (polymer/wall) interaction so that the solution and the final polymer film after the solvent is evaporated do not dewet the wall, and $\epsilon_W=1.0\epsilon$, $D=10\sigma$, $h_c=12\sigma$ for the NP/wall interaction. The upper wall is purely repulsive for all particles ($h_c=0.8583D$). For the NP systems, the starting state has 200 NPs randomly dispersed in a polymer solution that consists of 2.92 million solvent beads and 29,217 polymer chains of length 100, with $L_x = L_y =203.1\sigma$ and $L_z=300\sigma$. The volume fraction of the NPs is about $10\%$ and of the polymer is about $45\%$. We equilibrated the system for at least $2\times 10^4\tau$, with $\tau \equiv\sqrt{m\sigma^2/\epsilon}$ as the time unit, before evaporating the solvent. The equilibrium liquid/vapor interface is at $z \sim 200\sigma$. The equilibrium densities of solvent beads in the liquid and vapor phase are $\sim 0.35m/\sigma^{3}$ and $0.056m/\sigma^{3}$, respectively. The initial density of the polymer is $\sim 0.35m/\sigma^{3}$ as well. For the neat polymer solution, the number of solvent and polymer beads is 1 million each and the simulation box has dimensions $L_x = L_y =135.7\sigma$ and $L_z=215\sigma$ with the equilibrium liquid/vapor interface at $z \sim 135\sigma$. When the evaporation process of the solvent is initiated, the upper wall is moved to $L_z + 20\sigma$ with the region $[L_z, L_z + 20 \sigma]$ designated as the deletion zone. Every $0.5\tau$ all the solvent beads in the deletion zone are removed, effectively mimicking fast solvent evaporation into a vacuum. 

All simulations were performed using Large-scale Atomic/Molecular Massively Parallel Simulator (LAMMPS).\cite{plimpton95,lammps} The equations of motion are integrated using a velocity-Verlet algorithm with a time step $\delta t =0.005\tau$. During the equilibration, the temperature $T$ is held at $1.0\epsilon/k_{\rm B}$ by weakly coupling all beads to a Langevin thermostat with a damping constant $0.1\tau^{-1}$. Once the liquid/vapor interface is equilibrated, the Langevin thermostat is removed except for solvent and polymer beads within $10\sigma$ or $5\sigma$ of the lower wall for the systems with and without NPs, respectively.\cite{cheng11}

\section{Additional Simulation Results}

To confirm that the simulation results reported in the main text are robust with respect to the starting volume fraction of the NPs and polymer, we have modeled a system with the same number of 200 NPs and 29,217 linear polymer chains of length 100, but with the number of solvent beads increased to 8.9 million. The volume fraction of the NPs and polymer in the equilibrium system is about $5\%$ and $24\%$. During solvent evaporation all the phenomena are similar with those in the main text with higher initial volume fraction of the NPs and polymer. The surface polymer film quickly forms. After about $33\%$ of the solvent is evaporated, the exclusion of NPs from the surface polymer film for the weakly interaction case $A_{\rm np}=80\epsilon$ was observed as shown in Fig. \ref{more_solvent}. While a few NPs diffuse to the surface of the polymer film, most of the NPs are distributed in the interior of the film and accumulate in the region below the surface polymer layer.

\setcounter{figure}{0}
\makeatletter 
\renewcommand{\thefigure}{S\@arabic\c@figure}
\makeatother

\begin{figure}[ht]
\centering
\includegraphics[width=3.25in]{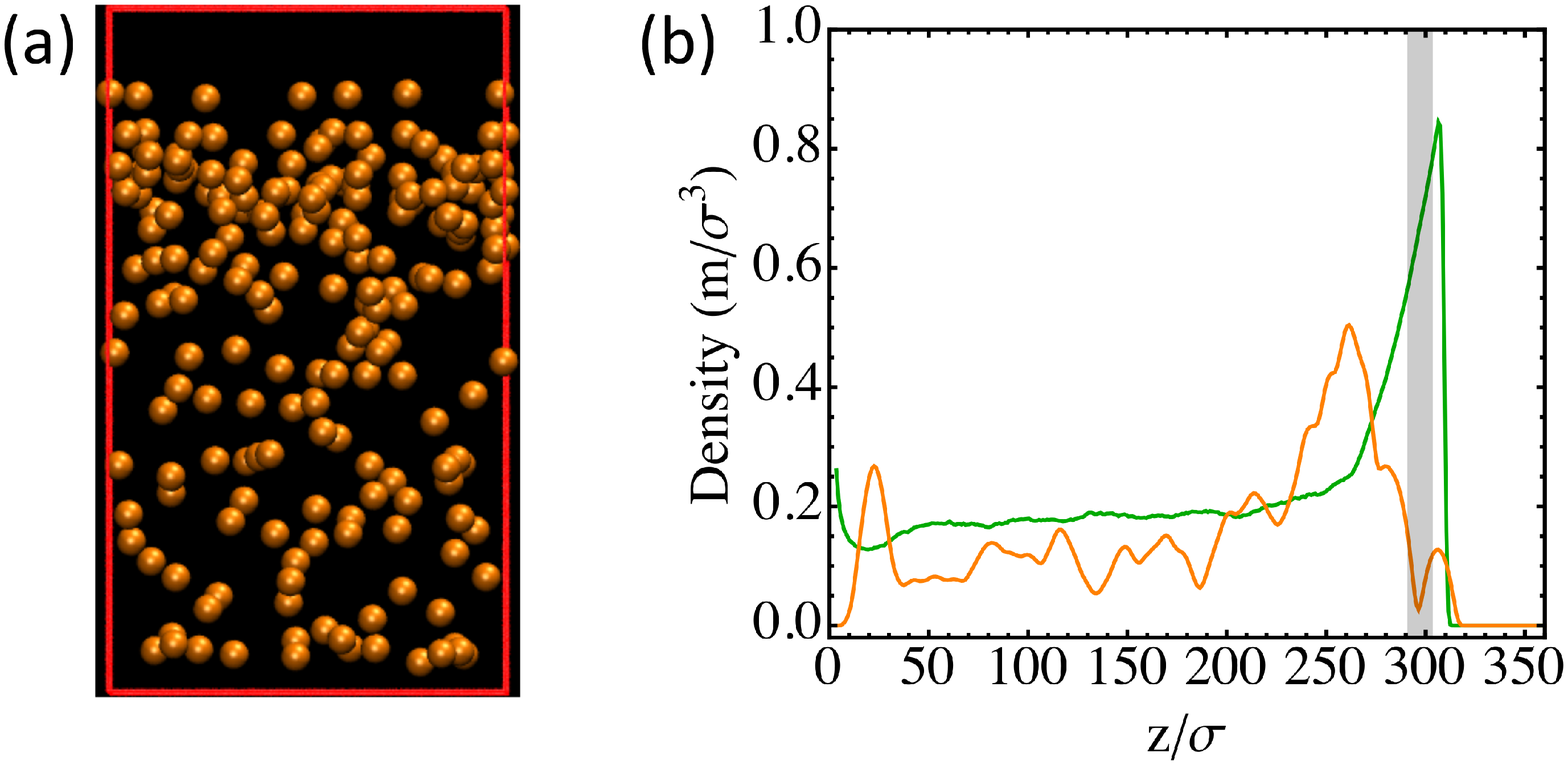}
\caption{(a) Snapshots of NP dispersion in the polymer film and (b) Density profiles of the polymer (green) and NPs (orange) after $\sim 33\%$ of the solvent is evaporated at $A_{\rm np}=80\epsilon$. The NPs are almost excluded in the region designated by the gray bar that is located in the surface polymer layer.}
\label{more_solvent}
\end{figure}

\end{document}